\documentclass[preprint2]{aastex}

\slugcomment{To appear in ApJ}

\shorttitle{Black hole neutron star mergers}
\shortauthors{S. Rosswog}

\begin{document}


\title{On the viability of neutron star black hole binaries as central engines
  of gamma-ray bursts}

\author{S. Rosswog\altaffilmark{1}}
\affil{School of Engineering and Science, International University
       Bremen, Germany}

\newcommand\Mesz{M\'esz\'aros~}
\newcommand\Pacz{Paczy\'nski~}
\newcommand\Kluz{Klu\'zniak~}
\newcommand\p{$e^\pm \;$}
\newcommand\msun{M$_{\odot}$}
\newcommand\Msun{M$_{\odot}$ }
\newcommand\be{\begin{equation}}
\newcommand\ee{\end{equation}}
\newcommand\bi{\begin{itemize}}
\newcommand\ei{\end{itemize}}
\newcommand\bea{\begin{eqnarray}}
\newcommand\eea{\end{eqnarray}}
\newcommand\gcc{g cm$^{-3}$}

\begin{abstract}

I discuss three-dimensional SPH simulations of neutron star black hole (BH)
encounters. The calculations are performed using a nuclear equation of state
and a multi-flavor neutrino treatment, general relativistic effects are
mimicked using the Paczynski-Wiita pseudo-potential and gravitational
radiation reaction forces.\\  
Most of the explored mass range (14 to 20 \msun) has not been
considered before in numerical simulations. The neutron star is always
disrupted during the first 
approach after most of its mass has been transferred directly into the
hole. In none of the analyzed cases episodic mass transfer is found. 
For the lower end of the mass range ($M_{\rm BH} \le 16$ \msun) an
accretion disk of moderate density ($\rho \sim 10^{10}$ \gcc) and temperature
($T < 2.5$ MeV) forms around the hole; the rest of the material forms a rapidly
expanding tidal tail, up to 0.2 \Msun of which are unbound. 
For higher mass black holes ($M_{\rm BH} \ge 18$ \msun) almost the complete
neutron star disappears in the hole without forming any accretion disk. In
these cases a small fraction of the star (between 0.01 and 0.08 \msun) is spun
up by gravitational torques and dynamically ejected.\\
None of the investigated systems of this study yields conditions that are
promising to launch a GRB. While we cannot completely exclude that a subset of
neutron 
star black hole binaries, maybe black holes with low masses and very large
initial spins, can produce a GRB, this seems to happen -if at all- only in a
restricted region of the available parameter space. I argue that the
difficulty to form promising disks together with the absence of any observed
neutron star black hole binary may mean that they are insignificant as central engines of
the observed, short-hard GRBs and that the vast majority of the latter ones is
caused by double neutron star coalescences.  
\end{abstract}

\keywords{black hole physics---hydrodynamics---methods: numerical---nuclear
  reactions, nucleosynthesis, abundances---gamma rays: bursts---}

\section{Introduction}
Neutron star binary systems have been recognized as potential central engines of
gamma-ray bursts (GRBs) already two decades ago, they have been mentioned in
Paczynski (1986) and Goodman et al. (1987) and discussed in more detail by
Eichler et al. (1989) (for a more complete bibliography we refer to existing reviews,
e.g. Meszaros 2002 or Piran 2005). Paczynski (1991) discussed systems
containing a neutron star (NS) and a stellar mass black hole (BH) as a possible GRB
engine. These days, compact binary systems are considered  the
'standard model' for the subclass of short gamma-ray bursts, that last
typically for about 0.3 s (Kouveliotou et al. 1993). While NSBH binaries are
usually just considered to be a minor variation on the topic of double neutron
star merger (DNS), it has been pointed out recently (Rosswog et al. 2004) that
it is not obvious, that such a coalescence will automatically produce a hot and
massive accretion disk around the hole. Therefore, its role for gamma-ray
bursts needs further investigations.\\ 
During the disruption process tidal torques are expected to eject material
into highly eccentric, possibly unbound orbits. This debris is extremely
neutron rich, $Y_e \sim 0.1$, and therefore (if ejected at appropriate rates) 
holds the promise to be one of the still much debated sources of r-process
elements (Lattimer and Schramm 1974 and 1976).\\  
Moreover, NSBH systems are generally considered promising sources for
ground-based gravitational wave detectors such as LIGO \citep{abramovici92},
GEO600 (Luck et al. 2001), VIRGO (Caron et al. 1997) and TAMA (Tagoshi et
al. 2001).\\ 
It is worth pointing out in this context that there is a controversy
about the rates at which NSBH mergers do occur. Bethe and Brown (1998)
argued that NSBH should merge about an order of magnitude more frequently than
DNS, while a recent study by Pfahl et al. (2005) comes to the conclusion that
the number of NSBH systems in the Galaxy should be below 1 \% of the number of
double neutron star systems. The current observational status is that
there are 8 observed DNS (Stairs 2004), while not a single NSBH binary has
been discovered yet.\\ 
Neutron star black hole merger simulations have been performed by several
groups. Janka et al. (1999) used a grid based hydrodynamics code together with
a nuclear equation of state (EOS) and a neutrino leakage scheme to explore the
role of these mergers in GRB context. Lee (2000, 2001) and Lee and Kluzniak
(1999a,b) used smoothed particle hydrodynamics with polytropic equations of
state to explore the sensitivity of the results on the adiabatic exponent of
the EOS. Lee and Ramirez-Ruiz (2002) have analyzed the flow pattern within an
accretion disk around a BH and, in a recent paper (2004), they included
neutrino processes in their simulations. Setiawan et al. (2004) constructed
disks around stellar mass black holes and followed their evolution including
explicit viscosity, Rosswog et al. (2004) investigated the dynamics of the
accretion process and the {\em formation} of a disk during the merger.\\ 
Stellar mass black holes formed in core collapse supernovae are thaught to be
born with masses ranging from about 3 to 20 \Msun (Fryer and Kalogera
2001). Numerical simulations have so far only explored black hole masses up to
14 \Msun  (Janka et al. 1999 and Lee and
Kluzniak 1999a,b and Lee 2000, 2001 used black holes up to 10 \msun,  Rosswog
et al. 2004 explored masses up to 14 \msun). In this paper we will focus on
black holes with masses ranging from 14 to 20 \msun. The 14 \msun case has
been explored previously (Rosswog et al. 2004) using the same microphysics but
a purely Newtonian BH-potential and may therefore serve to gauge the effect of the
Paczynski-Wiita pseudo-potential.\\
A simple estimate for the radius where a star around a BH is disrupted is the
tidal radius, R$_{\rm tid}= \left(\frac{{M}_{BH}}{M_{NS}}\right)^{1/3} R_{\rm
  NS}$. As it grows slower with the black hole mass than the gravitational
radius, it is expected that higher mass BHs will have more difficulties
building up massive disks. It has to be stressed, however, that these
simple estimates should only be used for qualitative statements as the
detailed numbers are off by factors of a few. Therefore such estimates
would lead to completely wrong predictions for the system dynamics.

\section{Simulations}

The calculations are performed with a 3D smoothed particle hydrodynamics code
that has been developed to simulate compact objects (for details see Rosswog
et al. 2000, Rosswog and Davies 2002, Rosswog and Liebend\"orfer 2003). 
Particular attention has been payed to avoid artifacts from the use of
artificial viscosity, away from shocks artificial viscosity is essentially
absent (Rosswog et al. 2000, Rosswog and Davies 2002).
The set of hydrodynamic equations is closed with a temperature and composition
dependent nuclear relativistic mean field equation of state based on the
tables of Shen et al. (1998a, 1998b) that has been smoothly extended to the
low-density regime (Rosswog and Davies 2002). Local cooling and compositional
changes due to weak interactions are accounted for with a multi-flavor
neutrino treatment (Rosswog and Liebend\"orfer 2003).\\ 
The self-gravity of the neutron star material is calculated using a binary
tree (e.g. Benz et al. 1990). 
The Paczynski-Wiita pseudo-potential is used to approximately take into
account the presence of general relativistic effects around the black hole
such as the presence of a last stable orbit. Clearly, the use of the
Paczynski-Wiita potential is not a 
complete substitute for fully fledged general relativistic hydrodynamic
simulation around a Schwarzschild black hole. It has, however, turned out to
be astonishingly accurate: it provides the exact values for the last stable
circular orbit ($R_{\rm isco}$= 6 M$_{\rm BH}$; $G=c=1$ throughout this paper)
and the marginally bound orbit (R$_{\rm mb}$= 4 M$_{\rm BH}$).  Direct 
comparisons with general relativistic solutions in a Schwarzschild space
time show that the pseudo potential is able to capture the essentials of
general relativity and can reproduce accretion disk structures to an accuracy
of better than 10 $\%$ (see e.g. Artemova et al. 1996).\\
I restrict myself to initially corotating systems as the corresponding
equilibrium configurations can be constructed accurately using the hydrocode
itself (see Rosswog et al. 2004). The simulations presented here use up to
3$\cdot 10^{6}$ SPH particles and therefore are currently the best resolved 
models of neutron star black hole encounters (Rosswog et al. (2004) used up to
10$^{6}$ particles, Lee (2001) used slightly more than 80000 particles).\\
The neutron star always has a mass of 1.4 \Msun and a radius of 16 km, the
black hole masses vary from 14 to 20 \Msun, see Table 1.
 
\section{Results}
\begin{figure}
{\includegraphics[angle=-90,scale=.3]{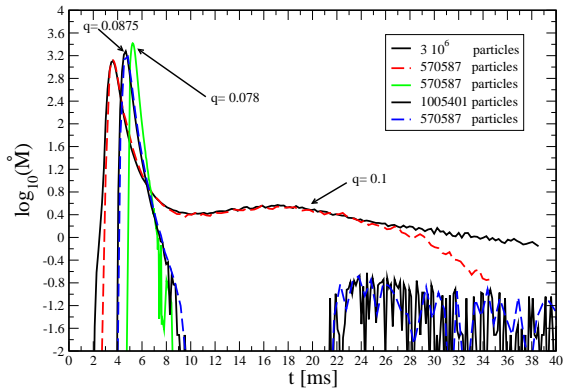}\includegraphics[angle=-90,scale=.3]{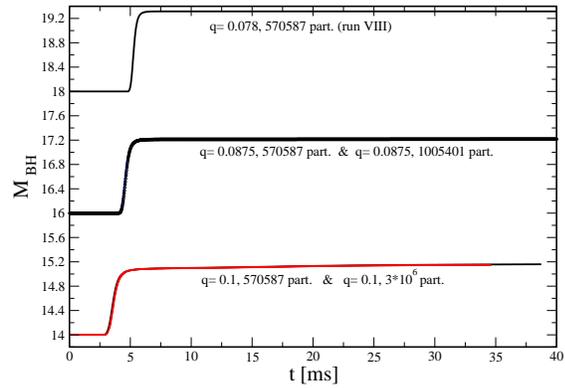}}
\caption{\label{mdot}
Left panel: The mass transfer rates as a function of time. In one case
(q=0.078, i.e M$_{\rm BH}$= 18 \msun) the mass transfer stops completely. This
is also true for the 20 \Msun case (not shown). 
Right panel: The growth of the black hole with time is shown for
  five of the runs.  Note that the runs that simulate the same
  systems (run I and II; run III and IV) with different resolutions yield
  nearly identical curves.} 
\end{figure}

\begin{figure}
{\includegraphics[angle=0,scale=.5]{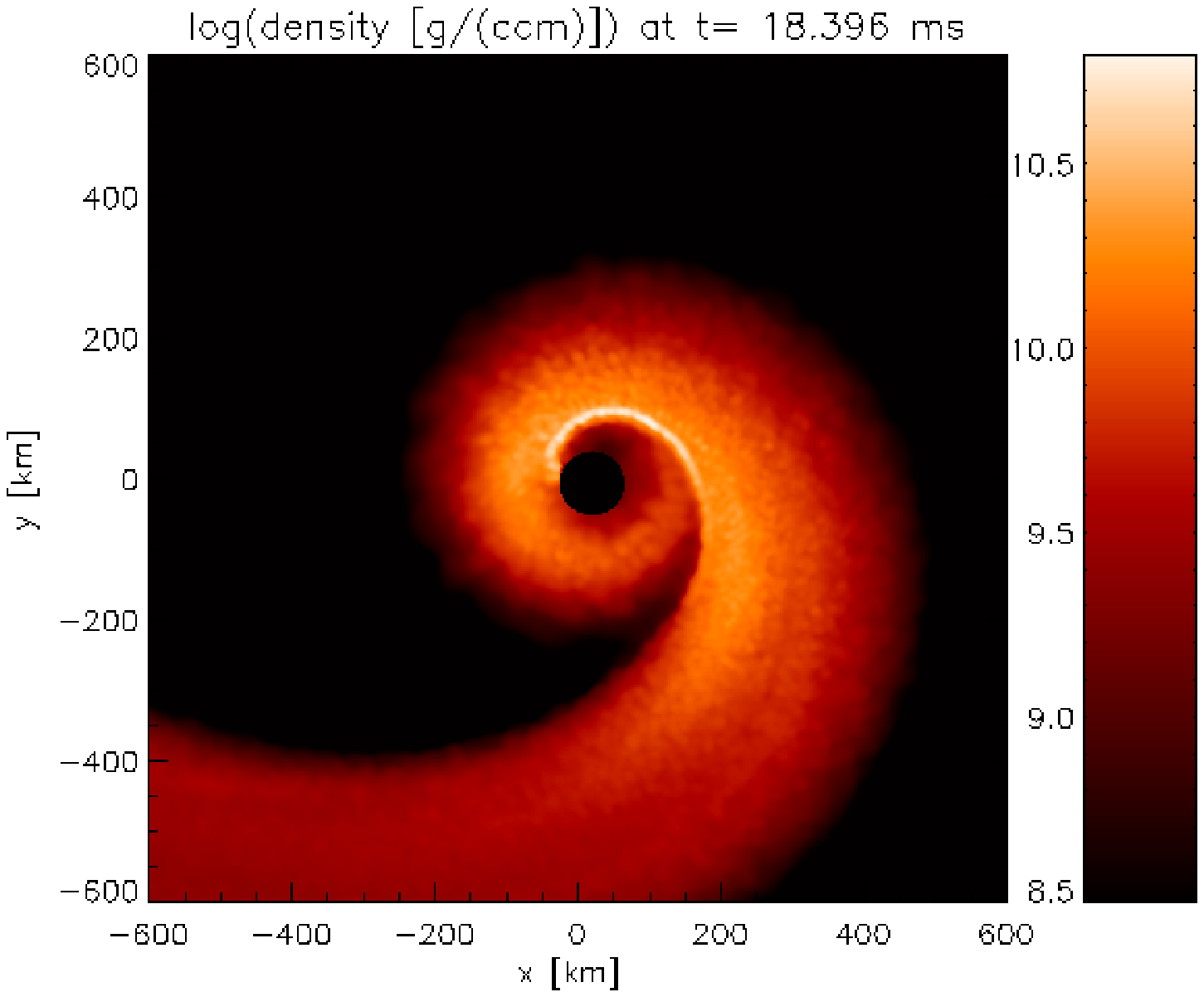}\includegraphics[angle=0,scale=.5]{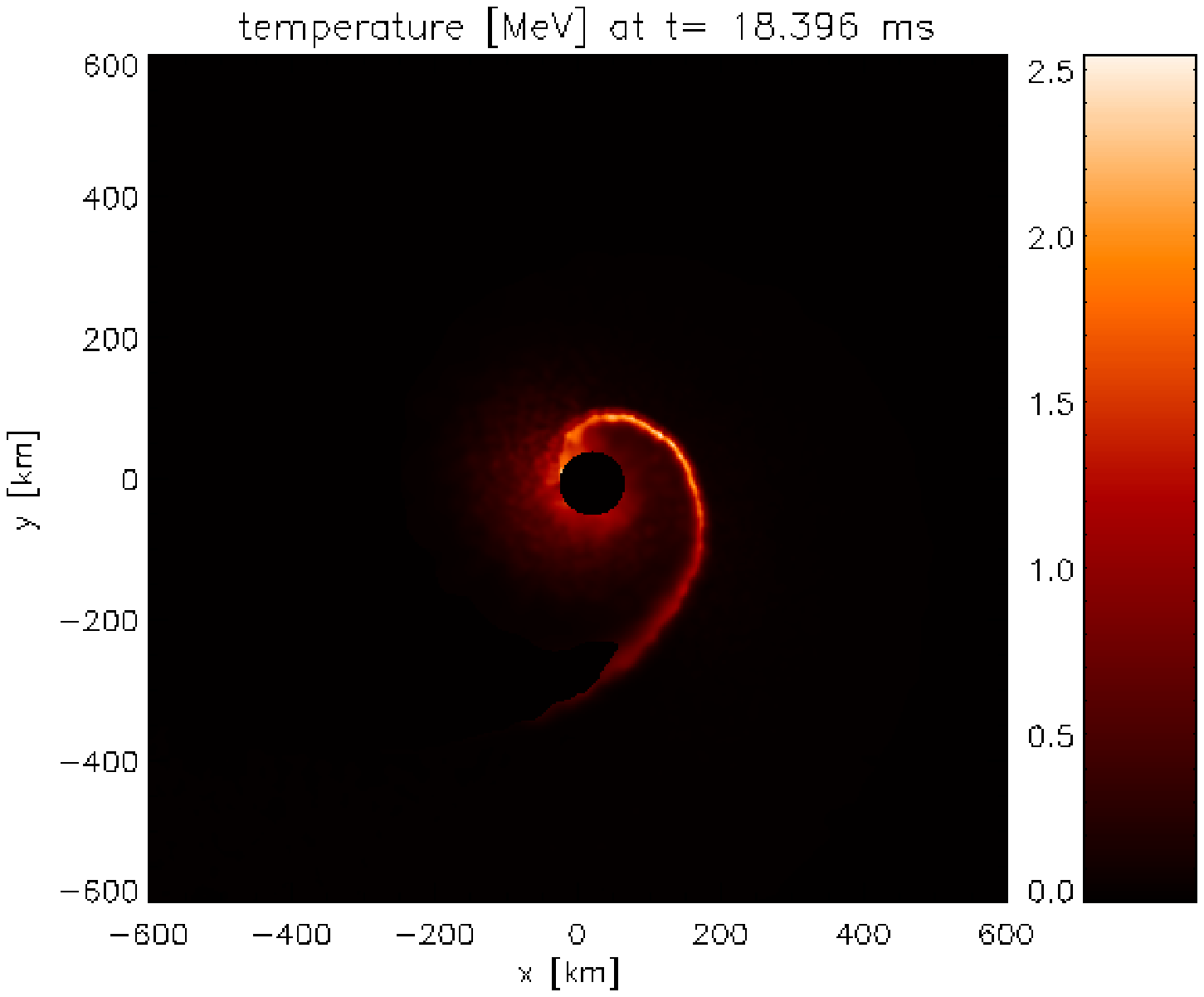}}

\caption{\label{rho_T_runII}Blow-up of the inner disk region of run II at
  t=18.396 ms after simulation start (left panel: log(density); right panel: temperature). Clearly visible are the shock, where the
  accretion stream interacts with itself and strong decrease in the density
  inside the last stable orbit.}
\end{figure}

The neutron star is always completely disrupted during the first encounter. In
all of the cases a large portion of the neutron star is transferred directly
into the hole without any accretion disk formation. The corresponding peak
accretion rates exceed 1000 \msun/s for about 1 ms, after this short episode
they drop by at least two orders of magnitude, see panel one in Figure
\ref{mdot}.\\ 
It is instructive to compare the 14 \Msun case to the corresponding case of
our previous study (Rosswog et al. 2004), where we had used a Newtonian BH
potential. In the purely Newtonian case we found episodic mass transfer with a
low-mass, ``mini neutron star'' surviving throughout the whole simulation or
about eight close encounters. In the  Paczynski-Wiita case about 1.15 \Msun
(see panel two in Figure \ref{mdot}) are transferred directly into the hole,
the rest forms a rapidly expanding tidal tail. The tidal tail still contains an
outward-moving density maximum (corresponding to the mini neutron star of the
Newtonian case), but its self-gravity is not strong enough to 
form a spherical object. The motion around the hole always has a strong
radial velocity component and is far from being Keplerian. The matter fraction
that is not swallowed during the first orbit collides with the accretion
stream forming a spiral shock (see Figure \ref{rho_T_runII}). In this shock
the temperatures slightly exceed 2 MeV, the other disk regions are
substantially colder. The disk is substantially diminished on timescales of
a few orbital periods. The densities never exceed $6\cdot10^{10}$ \gcc, the
neutrino luminosities reach peak values of only $2\cdot10^{50}$ erg/s and are
thus about three orders of magnitude lower than in our simulations of neutron
star binary mergers (see Rosswog and Liebend\"orfer 2003) where the same
microphysics was used.  The results are well converged, runs I and II show
excellent agreement in the BH masses and peak mass transfer rates. Some minor
deviations are visible at low mass transfer rates (see panel one in
Fig. \ref{mdot} and the distance, $R_{\rm MT}$, where numerically resolvable
mass transfer sets in; see column six in Table \ref{runs}).\\
The case with 16 \Msun BHs behaves qualitatively very similar to the 14 \Msun
BHs:  about 1.2 \Msun are transferred into the hole, the disk is slightly
less massive, hot and dense than the 14 \Msun case. Again, the two different
resolutions yield nearly identical results.\\ 
The systems containing BHs of 18 \Msun or more (runs V and VI) do not form
accretion disks at all. Almost the complete neutron star flows via the inner
Lagrange point directly into the hole, only a small fraction of the star is
spun up enough by tidal torques to be dynamically ejected, see last column in
Table \ref{runs}. In these cases the remnant consists of the black hole
(without any accretion disk) and a rapidly expanding, concentric (half-)ring of
neutron-rich debris material (0.08 \Msun for the 18  and 0.01 \Msun for 20
\Msun BH).\\
The ejected mass fraction found in these simulations is near the range
estimated by Lattimer and Schramm (1974 and 1976), they estimated
$0.05\pm0.05$ \msun. Our nucleosynthetic 
calculations for such debris material for the neutron star merger case
\citep{freiburg99} yielded excellent agreement with the observed r-process
abundances from around Barium up to beyond the platinum peak. If
large parts of the disrupted neutron star should form r-process material, 
a conflict with the observed element ratios in metal-poor halo stars might arise 
(Argast et al. 2004), if their coalescence rates are similar to those of double
 neutron stars (DNS). The problem is avoided if NSBH coalesce much less frequently 
than DNS. This would be consistent with the non-observation of any NSBH-system 
(currently 8 DNS systems are known, see 
Stairs 2004) and the result of recent studies
(Pfahl et al. 2005) that estimate their number in the Galactic disk to be less than
0.1-1\% of the number of DNS.\\ 
It is worth pointing out that in some of the investigated cases mass transfer
sets in only at distances considerably smaller than $R_{\rm isco}$ (e.g. run
VI, see Table \ref{runs}). This however, does not mean that the star is
swallowed as a whole (keep in mind that $R_{\rm isco}$ refers to the case
without self-gravity and to circular motion; both conditions are not satisfied
 here). In the 16 \Msun BH case still a tidal tail and accretion disk 
forms, in the higher mass cases at least some material escapes from
being drawn into the hole.

\section{Discussion}
We consider all the approximations made to be valid to a high degree. If an
accretion disk forms at all (i.e. for the BH masses at the lower end of the
explored range) it is of only moderate density ($\sim 10^{10}$ \gcc) and 
completely transparent to neutrinos. Therefore the neutrino emission
results cannot be influenced by our flux-limited diffusion
treatment. Moreover, the results are numerically converged, different
numerical resolution yields for the gross properties almost identical
results. The BHs are massive enough to dominate the space-time completely and as
they are spun up to spin parameters of only 0.2 (see Table 1), we consider the use of
PW-potentials a very good approximation (note that for a=0.2 the event horizon
moves from 2 to 1.98 and the last stable orbit from 6 to 5.33 gravitational
radii).\\ 
None of the investigated NSBH systems yields disks that are promising as GRB
engines. The disks formed in the low mass BH cases are relatively cold and of low
density, the neutrino luminosities are more than two orders of magnitude below our
results from the neutron star merger case (Rosswog and Liebend\"orfer, 2003).
In the latter case the GRBs launched via $\nu \bar{\nu}$-annihilation were
rather weak by GRB standards and, as the neutrino luminosities enter
quadratically in the energy deposition rate, $\nu
\bar{\nu}$-annihilation does not seem a viable GRB mechanism for the investigated
NSBH systems. Due to the lower densities the magnetic fields that can be
anchored in the disk are substantially lower that in the DNS case, whether
they still can launch a GRB remains to be explored in the future.
BHs beyond $\sim 18$ \Msun do not lead to any accretion disk formation at all
and can therefore be ruled out as sources of GRBs.\\
We still cannot  generally rule out NSBH-systems as central engines
of GRBs. One might speculate, for example, about about an extremly high disk viscosity.
Another possibility is the nuclear EOS. Our previous investigations (Rosswog et
al. 2004) showed that the dynamics of the merger is sensitive to the
EOS. Here, I used a relativistic mean field EOS with neutrons and protons as
the only baryonic constituents of matter. This EOS is certainly on the 
stiff side of the possible range of nuclear equations of state. Maybe a
substantially softer EOS could make the outcome of the merger more promising
to launch a GRB. The possibly most robust way out, however, are BHs that are
already from the beginning 
spinning very rapidly and that are spun up during the merger to values very
close to the maximum spin parameter of $a=1$. In this case both the position of
the last stable orbit and the event horizon move to 1 M$_{\rm BH}$ and thus
closer to the hole. Therefore, much higher temperatures and densities might be
reached in the inner disk regions.\\
It seems that a large part of the parameter space does not yield conditions that
are promising to launch GRBs. This difficulty to form promising disks together
with the absence of any observed NSBH system may mean that NSBH binaries are
insignificant as central engines of the observed, short-hard GRBs and that the
majority of the latter ones is caused by double neutron star coalescences.  
NSBH mergers may just manifest themselves as sources of gravitational waves and
transient X-rays.

\acknowledgments
It is a pleasure to thank Jim Lattimer, William Lee, Enrico Ramirez-Ruiz, Roland Speith and
Christophe Winisdoerffer for useful discussions and the INFN in Catania for
their hospitality. The calculations reported here have been performed on the
JUMP supercomputer of the H\"ochstleistungsrechenzentrum J\"ulich.




\clearpage


\begin{table}
\caption{Summary of the different runs. M$_{\rm BH}$: black hole mass;
  $q$=M$_{\rm NS}$/M$_{\rm BH}$; R$_{\rm tid}$: tidal radius;
$a_0:$ initial separation; R$_{\rm isco}$: last stable orbit Schwarzschild
black hole; R$_{MT}$: distance where numerically resolvable mass transfer sets
in; $\#$ part.: SPH particle number; $T_{\rm sim}:$ simulated duration;
a$_{\rm BH}$ is the dimensionless black hole spin parameter; M$_{\rm ej}$
refers to the material that is dynamically ejected during the merger.}
\begin{flushleft}
\hspace*{-1cm}\begin{tabular}{ccccrcccccccccc} \tableline \noalign{\smallskip}
run & M$_{\rm BH}$ [\msun]/$q$ & R$_{\rm tid}$ [km] & $a_{0}$ [km]    & R$_{\rm isco}$ [km]&
R$_{MT}$  [km] & \# part. & $T_{\rm sim}$ [ms]& a$_{\rm BH}$ ($T_{\rm sim}$)  & M$_{\rm ej}$ [\msun]\\ \tableline \\
%
%
I    & 14/0.1   &   36.1         &  127.5  &  124.1         &  117 &
570587  &    34.6    & 0.196  & 0.20  \\
%
%
%
%
II  &  14/0.1  &   36.1        &   127.5 &  124.1         & 125 & $3 \cdot
10^6$ &  40.8  & 0.200 & 0.20    \\
%
%
%
%
III    & 16/0.0875&   37.7        &  145.5  &  141.8         & 122 &
570587  &    78.1  &   0.197  & 0.15  \\
%
%
%
%
IV    & 16/0.0875&   37.7        &  145.5  &  141.8         & 123 &
1005401 &    60.9  & 0.197   & 0.15    \\
%
%
%
V    & 18/0.0778      &   39.3       &  162  &  159.6         & 123 &
570587  &   50.4   & 0.201  & 0.08   \\
%
%
%
%
VI    & 20/0.07      &   40.7      &  187.5  &  177.3        & 128  &
1503419  &   179.6 & 0.198  & 0.01    \\
\end{tabular}
\end{flushleft}
\label{runs}
\end{table} 

\clearpage

\clearpage


\end{document}